\documentclass[11pt, oneside]{article}

\usepackage[a4paper, total={6.2in, 9in}]{geometry}
\pdfoutput=1
\usepackage{graphicx}
\usepackage{amssymb}
\usepackage{xcolor}
\usepackage[T1]{fontenc}
\usepackage{lmodern}
\usepackage[normalem]{ulem} 

\usepackage{etoolbox}
\let\mybibitem\bibitem
\renewcommand{\bibitem}[1]{%
  \ifstrequal{#1}{nature}
    {\color{red}\mybibitem{#1}}
    {\color{black}\mybibitem{#1}}%
}

\PassOptionsToPackage{hyphens}{url}\usepackage[hidelinks]{hyperref}
\usepackage{fancyhdr}                           
\fancypagestyle{firstpage} 
{
   \fancyhf{}
   \fancyfoot[C]{\vspace{-12mm}\thepage}

   \fancyfoot[L]{\vspace{-5mm}\noindent\rule{6.2in}{0.4pt}
     \scriptsize{ 
         \textcopyright \hspace{0.5mm} Christopher D. Clack  2017-2018 \\
       This work is licensed under a \href{https://creativecommons.org/licenses/by/4.0/}{Creative Commons Attribution 4.0 International License (CC
       BY).}  \\ Provided you adhere to the CC BY license, including as to attribution, you are
       free to copy and redistribute this work in any medium or format and remix, transform,
       and build upon the work for any purpose, even commercially.                  \\
       \vspace{2mm}
       Electronic copy is available at \href{http://arxiv.org/abs/1711.10964}{http://arxiv.org/abs/1711.10964}\hspace{0.5cm}Contact the author at clack@cs.ucl.ac.uk
     }
 }
}

\title{\vspace{-1cm}Design discussion on the ISDA Common Domain Model}

\usepackage{anyfontsize}
\author{
  \begin{tabular}{c} 
    {\fontsize{9.75}{1cm}\selectfont Christopher D. Clack} \\ 
    {\fontsize{9.75}{1cm}\selectfont Centre for Blockchain Technologies} \\ 
    {\fontsize{9.75}{1cm}\selectfont Department of Computer Science} \\ 
    {\fontsize{9.75}{1cm}\selectfont University College London} 
  \end{tabular} 
}

\date{\small 29 November 2017\break(Revised 8 March 2018)} 

\begin{document}
\maketitle
\thispagestyle{firstpage} 
\vspace{-1cm}
\begin{abstract}
A new initiative from the International Swaps and Derivatives Association (ISDA) aims to establish a  ``Common Domain Model'' (ISDA CDM): a new standard for data and process representation across the full range of 
derivatives instruments.  
Design of the ISDA CDM is at an early stage and the draft definition contains considerable complexity.
This paper contributes by offering insight, analysis and discussion relating to key topics in the design space such as data lineage, timestamps, consistency, operations, events, state and state transitions.
\end{abstract}

\section{Introduction}
\label{sec:introduction}
There is considerable interest in ISDA's draft design for a Common Domain Model (ISDA CDM) \cite{ISDA-CDM}.
Although not yet in final form, the ambition is clear: to create a common representation for a wide range of derivatives instruments so that substantial benefits may accrue from standardisation.  
The ISDA CDM draft definition is a complex document, for example with definitions in a mathematical style appearing almost from the start, and there is a need to provide commentary, discussion and feedback.  There is also a need to provide rigour and additional thinking to contribute to improvements in the next iteration of the definition.

This paper aims to help people who are looking to understand the ISDA CDM draft definition, 
and to support those aiming to undertake design and prototyping of the ISDA CDM in software.  
There are many topics that could be covered in this paper, and to provide the most value initial focus is given to the following topics that are considered foundational: data lineage (the ability to track the antecedents of derived data), timestamps, consistency (of replicated data in distributed implementations), operations and events (since these are central to the ISDA CDM definition), and the concepts of state and state transition (which require further rigorous analysis).

The summary includes a list of the more important issues that arise during the analysis and discussion in this paper, and the appendix provides a list of features and benefits culled from  \cite{ISDA-CDM-2} with the expectation that further analysis can repurpose these benefits as principles, to provide a more rigorous context to guide and motivate design choices.

The paper will be of interest to business architects, technology architects, specialists in derivatives, regulators and academics.  Issues and concepts are discussed using reasonably straightforward language, though in places it is necessary to engage with technical detail and precision in specifications.  This is a working document: it is hoped that the issues and views raised in this paper will stimulate debate and the author looks forward to receiving feedback to help improve the paper.

\section{ISDA CDM engagement and technology platform}
\label{sec:rollout}
The benefits that will accrue to the derivatives community as a result of the ISDA CDM are likely to be directly linked to the scale of engagement across a range of different dimensions:
\begin{itemize}
\item
the number of participants,
\item
the type of participants,
\item
the number of products, and
\item
the degree of integration throughout end-to-end processes in the full product lifecycle.
\end{itemize}

\noindent
It will be important to encourage full engagement with the ISDA CDM, rather than for example merely the writing of adapters for existing data structures and processes, and it will be important to get engagement from technology vendors to support participants in integrating the ISDA CDM into their end to end processes.  Having multiple financial market infrastructures (such as multiple clearing houses) support the ISDA CDM will also be crucial, since there will be a supra-linear relationship between the degree of engagement across the derivatives community and the consequential benefits experienced by all.

A  ``big bang'' approach where all participants engage with the ISDA CDM simultaneously is highly unlikely to occur.  Rather, it must be expected that engagement will be gradual, with an increasing number of participants engaging to an increasing extent as time progresses.

In order to encourage rapid and broad adoption of the ISDA CDM, an important issue for discussion is the extent to
which it may enforce use of any particular technology platforms.
The draft definition \cite{ISDA-CDM}  embraces technology-independence when it states ``While the model definition is generic, and could be adopted via any technology...'', yet there is also an understandable desire to encourage novel technology and the quote from the definition continues ``... the implementation is targeted at distributed ledgers ...''.   This overstates the case:  there is very little (arguably nothing) in the draft definition that is {\em targeted} at distributed ledger technology (DLT), nor is there any aspect of the definition that {\em requires} DLT.  The quote ends "...to exploit the embedded Lineage and Consistency properties of distributed ledgers'', yet as is argued later in this paper Lineage can be provided by any implementation and Consistency is not an issue if a non-distributed implementation is chosen.   This, then, is good: there is nothing in the draft definition to preclude implementation on a particular technology platform, and this strengthens the ISDA CDM's position as a common standard.

An implementor of the ISDA CDM may choose to use either a centralised platform or a distributed platform.  If a distributed architecture is chosen, a further choice is whether or not to use DLT (and why).  Finally, if DLT is chosen, there is a further choice of which kind of DLT to use.  The ISDA CDM is currently agnostic to these issues, and preferably will remain agnostic since this supports the broadest range of implementation options.  For example, a clearing house might choose to offer an interface to members based on the ISDA CDM data structures and processes, but where the implementation is internal to the clearing house and is based on a centralised architecture: the service operator would commit to be the provider of the authoritative central record (the ``point of truth'') for its members.  And where a DLT implementation is available, several modes of engagement may be imagined:  (i) each participant might host its own DLT node; (ii) a participant's DLT node might be hosted by a third party, with participant access to the node; or (iii) existing interfaces might be used so that a participant does not need to be aware that DLT is being used.  The choice of engagement model may be determined by a participant's desire or capability to implement a DLT node, and no doubt many other ways of engaging with the ISDA CDM may be imagined.

\section{Abstraction Layer}
The ISDA CDM's common representation of data structures and processes can be conceptualised as an abstraction layer that provides a common standard for all supported products.  The ISDA CDM can be implemented using a number of different low-level technology platforms, and software vendors may also provide a range of high-level analysis and workflow applications.  The aim of the ISDA CDM to be a common standard is important because it can free up innovation in both low-level technology platforms (e.g. encouraging the use of DLT and smart contracts) and high-level applications (e.g. encouraging new analysis that leverages access to standardised data across products):  the aspiration is that providing a standard across all derivatives products at a middle level will create a ``plug and play'' infrastructure. 

A natural question therefore is ``what is the appropriate level at which this standard should be defined?'' --- i.e. should it be defined at a very high level that represents business processes easily yet is far removed from the details of the technology processes, or should it be defined at a very low level that is closely linked to the technology processes yet is far removed from the business processes?  Another way of asking this question would be to query whether the current draft definition of ISDA CDM operations and events is too generic or insufficiently generic.

There will be a ``Goldilocks'' level that is neither too high nor too low (neither too generic nor insufficiently generic), and which provides optimal benefits.  Whether the current ISDA CDM definition is currently at that ``Goldilocks'' level is a question of great practical relevance; at present it is not yet clear how to determine the ideal level, and no doubt this will benefit from further discussion.

The draft definition of the ISDA CDM is currently specified at a fairly low level, yet utilises a design pattern that introduces a degree of genericity:  in some respects this might be viewed as an ``abstract machine''.\footnote{Since an implementation is assumed, we might also use the term ``virtual machine''.}  This is advantageous in that it can be independent of the underlying technology platform, but it requires a complete and rigorous definition to ensure the abstract machine is correct, and as yet the draft definition does not provide this level of rigour.

\section{Data lineage}
Data lineage is a key desirable property that can be provided in many ways.  For example, the draft definition \cite{ISDA-CDM} focuses on a DLT implementation, yet data lineage can also be provided by a centralised implementation, or be achieved programmatically as illustrated below.

Data lineage facilitates analysis of data and of the functions that operate on data, such that for any data (whether an original observation, or a value derived from the operation of a function on some other values) it is possible to elucidate the provenance of that data --- i.e. where it came from, what other data was used in its formation, and which functions have been used in its calculation.

The property of data lineage can either be built-in to an implementation, or built-in at the level of an abstract machine, or within a compiler, or can be programmed explicitly as illustrated in the detailed example below.  

\subsection{Example of programmed data lineage}

This subsection provides a simple example of how data lineage can be programmed explicitly, as a way to illustrate the concept.  The examples are given as specifications, to make the explanations accessible to non-programmers (though the specifications are necessarily technical), yet the expectation is that programmers will immediately see how these specifications could be encoded.\footnote{The specifications use a syntax that is inspired by the functional programming language Miranda$^{TM}$ \cite{Turner1986,Thompson1995,Clack1995} because it has a parsimonious syntax that is well suited to specifications, yet is also executable \cite{Turner-executablespec}.  Miranda$^{TM}$ is a trademark of Research Software Limited.}

The specifications are given as {\em type specifications} for data and {\em function definitions}, where the latter illustrate how the lineage property is derived and constructed.  The most basic type for all data is assumed (for simplicity) to be numeric, indicated by the type name $num$.  This numeric data is then augmented with extra information about its provenance (where it has come from).  The combination of basic data and provenance information can be treated as a new user-defined type; this can be specified using the specification $newtypename\;::=\;Tag\;\;types\_of\_component\_data$.  As an example, consider the type specification for a new type called $augmentednum$ that will contain both a number (type $num$) and some provenance information (type $provenance$):
\[
augmentednum \;\;::= \;\;Aug \;\;num \;\;provenance
\]

\noindent
In the above specification, $Aug$ is the tag, and it is followed by two items giving (i) the type of the data ($num$) and (ii) the type of the additional information --- in this case the type name $provenance$, which is another new user-defined type as specified below:

\[
provenance \; \; ::= \; \; Orig \; text \; \; | \; \; Derived \; text \; [provenance]
\]

\noindent
The above specification uses a vertical bar ``$|$'' to indicate alternatives:  the specification states that anything of type $provenance$ will be {\bf either} (i) an originally observed number (using the tag $Orig$), for which 
a further description is provided of where the data was observed using information of type $text$, or (ii) a derived number (using the tag $Derived$), for which 
a further description (of type $text$) is provided of the function that was used to create the derived number followed by a ``list'' (designated by the use of square brackets: $[$ and $]$) of $provenance$ information about all of the data inputs to that function. A ``list'' is simply an ordered sequence of items.

Notice how $provenance$ is defined recursively (the name appears in its own definition).   This means that if the inputs to a function are themselves derived values then the available provenance information can contain the entire history of all values and how they were created, and this can be traced all the way back to the originally observed values (see examples below).

So far only type specifications have been given.  It is now necessary to give specifications for functions that take in numeric data and produce numeric data.  
Here is a very simple example of a function called $replace$ that takes two arguments (of type $augmentednum$ and $num$) and returns an augmented number with a new value:

\[
replace \;((Aug \; x \;p), y) = Aug \;y \;(Derived \; ''replace'' \;[p])
\]

\noindent
The above specification defines $replace$: it takes two arguments, the first of which must be the tag $Aug$ followed by some data (given the name $x$) and some provenance information (given the name $p$), and the second of which is given the name $y$; and it returns a value of type $augmentednum$ comprising the tag $Aug$ followed by the value $y$ and the updated provenance information (using the tag $Derived$ followed by text to say which function was used, followed by a list of provenance information that contains the single item $p$ from the input).

Here is a second example of a function called $aplus$ that adds together two augmented values and returns the result as an augmented value:

\[
aplus \; ((Aug \;x \;p1),\; (Aug \;y \;p2)) \;= \;Aug \;(x+y) \;(Derived \;''aplus''\; [p1, p2])
\]

\noindent
The function $aplus$ takes in two values $(Aug \;x \;p1)$ and $(Aug \;y \;p2)$ both of which contain both a number and some provenance information.  The function produces as its result the value $Aug \;(x+y) \;(Derived\; "aplus" \;[p1,p2])$ --- i.e. the output value is simply the result of the addition $x+y$ and the output provenance information is the tag $Derived$ followed by a string saying which function has just been used to create the output and followed by the provenance information from both inputs collected into a list $[p1, p2]$.

The operation of lineage can be demonstrated by applying the function $aplus$ to some actual data (where the data $ob1$ is first to be observed, $ob2$ is second to be observed, and $ob3$ is last to be observed, and they are added in time order):

\[
aplus \;((Aug \;3 \;(Orig \;''ob3'')),\; (aplus \;((Aug \;4 \;(Orig \;''ob2'')),\; (Aug \;5 \;(Orig \;''ob1'')))))
\]
 
\noindent
When this expression is evaluated, the following output is produced:
 
\[
Aug \;12 \;(Derived \;''aplus''\; [Orig \;''ob3'',\;Derived \;''aplus''\; [Orig \;''ob2'',\; Orig \;''ob1'']])
\]

\noindent
The above example illustrates that any function that subsequently uses the above result can access not only the numeric result $12$ but also the detailed provenance of what original data was used and which functions processed the data step by step.  If all functions were to operate on and return augmented numbers, the full provenance would always be available.

\subsection*{Storing historical values}
It is also possible to store a history of previous values inside the provenance.  This can be achieved programmatically by defining a new provenance type called $provenance1$ and a new augmented number type called $augmentednum1$ as follows:
\vspace{11pt}

\begin{tabular}{rl}
$augmentednum1 \; $&$::= \; Aug1 \; num \; provenance1$\\
$provenance1 \; $&$::= \; Orig1 \; text \;num \; | \;Derived1 \;text \;num \;[provenance1]$
\end{tabular}
\vspace{11pt}

\noindent
The functions $replace$ and $aplus$ can have history-saving variants called $replace1$ and $aplus1$ as follows:
\vspace{11pt}

\begin{tabular}{rl}
$replace1 \;((Aug1 \;x \;p1), \;y) $&$= \;Aug1 \;y \;(Derived1 \;''replace1''\; y \;[p1])$\\
$aplus1 \;((Aug1 \;x \;p1),\; (Aug1 \;y \;p2)) $&$= \;Aug1 \;(x+y) \;(Derived1 \;''aplus1''\; (x+y) \;[p1, p2])$
\end{tabular}
\vspace{11pt}

\noindent
The operation of history-saving lineage can be demonstrated by applying the function $aplus1$ to actual data as illustrated below, where the data is a variant of that used in the previous example of $aplus$ (the expression has been split across two lines for clarity):
\vspace{11pt}

\begin{tabular}{l}
$aplus1 \;((Aug1 \;3 \;(Orig1 \;''datum3''\; 3)), $\\
$      \;\;\;\;\;\;\;\;\;\;\;\;\; (aplus1 \;((Aug1 \;4 \;(Orig1 \;''datum2''\; 4)), \;(Aug1 \;5 \;(Orig1 \;''datum1''\; 5)))))$
\end{tabular}
 \vspace{11pt}

\noindent
When this is evaluated, the following output is produced (split across three lines for clarity):
\vspace{11pt}

\begin{tabular}{lll}
$Aug1 \;12 \;(Derived1 \;''aplus1''\; 12 $&$[Orig1 \;''datum3''\; 3, $\\
&$\;Derived1 \;''aplus1''\; 9$&$[Orig1 \;''datum2''\; 4, $\\
&&$\;Orig1 \;''datum1''\; 5]]) $
\end{tabular}

\subsection{Simple example of lineage in state transition}
The functions $replace$ and $replace1$ illustrate how an augmented value can be updated while tracking provenance.  An obvious question is how to extend this to updating a state, such that it could be incorporated into a  state transition system such as the ISDA CDM proposes.  

A later section discusses the notion of state in the draft definition of the ISDA CDM, and argues that the current state of a contract is not yet well defined in the draft definition.  There is no clear specification of the state items nor how they should be represented.  The following example therefore illustrates programmatic data lineage for a generic state that is an ordered sequence of (key, value) pairs.  The state is presumably queried from time to time to discover the value associated with a key (assumed to be a name), and 
presumably updated from time to time either to insert a new (key, value) pair or to change the value for an existing key.

This can be modelled with type and function specifications, as used above.
In the following example the new type $state$ is specified as a list (i.e. an ordered sequence of items, signified by the square brackets) of pairs (signified by the round brackets) each of which will contain a key of type $text$ and a value of type $augmentednum$.  

\[
state \;::=\;State \; [ \;(text,  augmentednum) \;]
\]

\noindent
Now all that is needed is a function to update this state.  For example, the function $update$ will take two arguments (a $state$, and a (key,value) pair of type $(text, augmentednum)$) and return an updated $state$:
\vspace{7pt}

\begin{tabular}{ll}
$update \;(State\;[\;] ,$&$(n,v))\;=\;State\; [(n, v)] $\\
$update \;(State ((k,av):st),$&$(n,v))$\\
&\hspace{-1.5cm}$= \;State\;((k, areplace \;(av,v)):st), \;if \;k=n$ \\
                             &\hspace{-1.5cm}$= \;prefix \;((k,av),\;update \;(State\;st, \;(n, v))), \;otherwise$\\
                             &\hspace{-1.5cm}$\;\;\;\;\;where$\\
                             &\hspace{-1.5cm}$\;\;\;\;\;prefix \;((x,y), \;(State \;z)) = State \;((x,y):z)$\\
                             &\hspace{-1.5cm}$\;\;\;\;\;areplace \;(Aug \;x \;p, \;Aug \;y \;q) = Aug \;y \;(Derived \;``areplace'' \;[p,q])$
\end{tabular}
\vspace{4pt}

\noindent
In the above definition there are two alternative definitions for {\em update} depending on whether the first argument of type {\em state} contains any (key,value) pairs. If it has no pairs (the empty list is signified by $[\;]$), the function {\em update}  returns a newly constructed value of type {\em state} containing the key from the second argument and the value from the second argument. If the first argument does contain at least one (key, value) pair, the second definition for {\em update}  is used: this checks the key in the first pair in the state's list to see whether it is the same as the key in the second argument. If a match is found, the resulting state is created by updating the value in the first pair of the state's list.  If a match is not found, the first pair in the list is added onto the result of a recursive call to the {\em update}  function (this has the effect of iterating down the list of pairs until a match is found, or until the {\em update}  function is called on a state with an empty list of pairs).  Two subsidiary functions are used, following the keyword ``where'': {\em areplace} is similar to {\em replace} but takes two augmented numbers as arguments, and {\em prefix} takes a pair and a state and adds the pair to the start of the state's list of pairs.

\noindent
If it were required to save a history of values in the provenance, the specification would look like this:
\vspace{11pt}

\begin{tabular}{ll}
$state1 \;::=\;State1 \; [ \;(text,\;augmentednum1) \;]$\\
\\
$update1 \;(State1\;[\;],\;\;\;\;\;\;\;\;\;\;\;\;\;\;\;\;(n,v)) \;=\;State1\; [(n, \;v)] $\\
$update1 \;(State1 ((k,av):st),\;(n,v)) $\\
&\hspace{-7.5cm}$= \;State1\;((k, (areplace1 \;(av,v))):st), \;if \;k=n$ \\
                             &\hspace{-7.5cm}$= \;prefix1 \;((k,av),\;(update1 \;((State1\;st), \;(n, v)))), \;otherwise$\\
                             &\hspace{-7.5cm}$\;\;\;\;\;where$\\
                             &\hspace{-7.5cm}$\;\;\;\;\;areplace1 \;(Aug1 \;x \;p, \;Aug1 \;y \;q) = Aug1 \;y\;(Derived1\;``areplace1''\;y\;[p,\;q])$\\
                                                          &\hspace{-7.5cm}$\;\;\;\;\;prefix1 \;((x,y), \;(State1 \;z)) = State1 \;((x,y):z)$
\end{tabular}
\vspace{11pt}

\subsection{More complex lineage issues}

\subsubsection*{Provenance trees}

The simple examples above only considered a function $replace$ that took in one argument, and a function $aplus$ that took in two arguments.  In the former case the provenance information was in the form of a linear chain: in the latter case the provenance information was in the form of a binary tree.  The ``binary tree'' is explained as follows:

\begin{itemize}
\item
In the definition of $provenance$, the $Derived$ tag can be followed by a textual description and then a list of items each of type $provenance$ (the list --- or sequence --- is signified by the square brackets in the type $[provenance]$):
\[
provenance \; \; ::= \; \; Orig \; text \; \; | \; \; Derived \; text \; [provenance]
\]
\item
A list is a collection of data items in the form of a chain (it is possible to precisely associate each item with a single integer to show its position on the chain).
\item
However, items are only added onto this chain by the function $aplus$, and this function only creates two items on the provenance list:
\[
aplus \; ((Aug \;x \;p1),\; (Aug \;y \;p2)) \;= \;Aug \;(x+y) \;(Derived \;''aplus''\; [p1, p2])
\]
\item
Of the two provenance items $p1$ and $p2$ placed on to the chain, either none, one, or both of them will have been created by a previous invocation of the $aplus$ function.  Thus there are four possibilities for $[p1,p2]$:
\vspace{11pt}

\begin{tabular}{ll}
$[Orig \;o1, $&$Orig \;o2]$\\
$[Derived \;''aplus''\; [q1, q2], $&$Orig \;o2]$\\
$[Orig \;o1, $&$Derived \;''aplus''\; [q3, q4]]$\\
$[Derived \;''aplus''\; [q1,q2], $&$Derived \;''aplus''\; [q3,q4]]$
\end{tabular}
\vspace{11pt}

In the above example, $o1$ and $o2$ are data descriptions, and $q1, q2, q3, q4$ are provenance information items deriving from other input data to other invocations of $aplus$.
\item
Thus, if we consider $Orig$ to be the leaf of a tree and $Derived$ to be a node of a tree with two sub-trees being the items $p1$ and $p2$ inside the list $[p1,p2]$, then we can say that a tree structure is embedded inside the list.  That is, we say that 
\[
Derived \;''aplus''\; [Orig \;''ob3'',\;Derived \;''aplus''\; [Orig \;''ob1'', \;Orig \;''ob2'']]
\]
\noindent
is a top tree node $Derived$ with a left descendant that is just the leaf $Orig \;''ob3''$ and with a right descendant that is a node $Derived$.  This second node itself has two descendants, which are the leaf $Orig \;''ob1''$ and the leaf $Orig \;''ob2''$.
\end{itemize}

\noindent
To be more precise, if in our specification all functions that returned something of type $augmentednum$ only took two arguments of type $augmentednum$ (no more, and no less) then every occurrence of the tag $Derived$ would be followed by a description and a provenance list {\bf with only two items in the list}, and this can be visualised as a tree structure where internal nodes all have exactly two sub-trees.  This tree structure is the provenance hierarchy.

However, in general, a function might take in any number of arguments of type $augmentednum$.  Here is an example of a function that takes in three values and adds them:
\vspace{11pt}

\begin{tabular}{ll}
$add3 \;((Aug \;x \;p1),\; (Aug \;y \;p2),\; (Aug \;z \;p3))\; = \;Aug\; (x+y+z) \;(Derived \;''add3''\; [p1,p2,p3])$
\end{tabular}
\vspace{11pt}

\noindent
If the function $add3$ were utilised, it would lead to a node in the provenance hierarchy with three sub-trees 
$[p1,p2,p3]$, and this would happen at each occasion the function was used.

\subsubsection*{Lineage information and standardised access}
As illustrated in the previous section, data lineage is in essence an information concept.  This explains why it is possible to capture the property of lineage at a variety of levels.  It has been demonstrated above that data lineage can be expressed programmatically and explicitly collected and organised by a programmer, but if it were captured in the definition of the ISDA CDM (at the level of the abstract machine) then it would be available to all software that accesses the ISDA CDM data representations and processes.  This would also provide the possibility of higher resolution lineage information, or lower resolution lineage information, according to the precise ISDA CDM specification.

For example, provenance information could be collected at each state-to-state transition for the ISDA CDM, or possibly at the start and end of common sequences of state-to-state transitions.  
In both cases, the ISDA CDM state would be augmented to include the provenance information, but the updating of provenance information would be achieved differently:
\begin{itemize}
\item
for the former case, provenance update could be built in to every ISDA CDM state transition, whereas 
\item
for the latter case there could be an explicit state transition that updates the provenance information 
and which could be called at the start and end of a sequence of state transitions (and the ISDA CDM specification could prescribe exactly when these provenance-update transitions must occur).
\end{itemize}

\noindent
It would of course be possible to defer definition of the data lineage data structures to the technology platform, but that might leave it open to the technology implementors to provide lineage in different ways, and this could work against the requirements for standardisation (and therefore make it more difficult to write standardised analysis software that accesses the lineage information).  To avoid loss of standardisation, it would be necessary to make precise specifications for the lineage data structures and how they would be updated and accessed --- expressed as constraints on the implementation.

Although this clearly requires further analysis, it appears preferable to embed the lineage property (and specification of the data formats) into the definition of the ISDA CDM, as described above, in such a way that efficient implementation is possible.

\section{Timestamps}

It is possible to extend the provenance information to include the precise date and time at which each original observation of data was made (e.g. the time at which a market rate was observed) and at which each function operated to create a derived instance.  This could be achieved trivially by embedding the date inside the textual description, but more usefully it could be held separately:

\[
provenance2 \; \; ::= \; \; Orig2 \; date\;text \; \; | \; \; Derived2 \; date\;text \; [provenance]
\]

\noindent
This notion of timestamping could be extended to other events (not just observation of data).  However, 
timestamps must be approached with care:

\begin{itemize}
\item
Is the precise time required, or is it the ordering of events that is required?
\item
If the implementation is non-distributed, then taking timestamps from the system clock may be sufficient.  
\item
If time is required to high precision, then that precision should be specified (since it may require investment in specialist technology).
\item
If the implementation is distributed across several computers, the system clocks of those computers may not be synchronised and it may be difficult to compare times safely.   There are known solutions to the clock synchronisation problem, but there is always a margin of error --- preferably the acceptable error margin will be specified.
\item
Furthermore, in a distributed implementation events may happen simultaneously: in which case how shall ``precise ordering'' be defined?  Is it sufficient to define a partial ordering rather than a total ordering on events? Existing solutions that should be considered include Lamport timestamps \cite{Lamport1978} and vector clocks \cite{Fidge1988,Mattern1988}.
\item
In a DLT implementation information about the events in the distributed system might be logged in a replicated (synchronised) data structure.  This could feasibly provide a precise ordering of which event information was added to the data structure before which others.  However:
\begin{itemize}
\item
delays and/or race hazards might occasionally lead to earlier event information being recorded after that for a later event;  
\item
the additions to the central logging data structure may happen as aggregated information with a much lower time resolution than is available from a system clock; 
\item
the time at which information about an event is added to the logging data structure might be considerably later than
time at which the event occurred; and 
\item
if precise time (rather than ordering) is required, it is likely to still be necessary to synchronise the clocks of all the computers  in the distributed ledger.
\end{itemize}
\end{itemize}

\noindent
If lineage is implemented as a provenance hierarchy built from the bottom up (as explained in the previous section) then that provenance hierarchy provides some sequencing information ``for free''.  This is because the items lower down in the tree must have occurred before the nodes higher up the tree were constructed.  The provenance hierarchy does not provide information about sequencing of two items at the same level in the provenance hierarchy, but timestamps could then be used to determine ordering.  It may also be possible to include further time sequencing information alongside a timestamp, to indicate known ordering of events even where they are not related via data dependency (and therefore may not have an obvious ordering within the provenance hierarchy).

If timestamps are important, then an important discussion will be whether they should be incorporated into the definition of the ISDA CDM rather than left to the whim of the implementation.  However, this is a nuanced discussion (as hinted at in the issues outlined above).

\section{Consistency}
\label{sec:consistency}
Consistency is a somewhat over-used word that can have different definitions according to context.  For example, in the context of database systems consistency is often defined to be the property that any transaction will move the database from one valid state to another valid state.  In the context of the ISDA CDM the term consistency is used to describe a property of a distributed implementation whereby multiple copies of the same version of an object should be consistent.

However, consistency can normally only be guaranteed within a defined timeframe --- there will always be delays in synchronising copies of objects after one copy has been updated.  Furthermore, the notion of consistency must be linked to versioning, since some version updates might not yet be synchronised, and conflict resolution must be performed if two conflicting updates are performed at the same time.  The definition of consistency should also consider issues of availability and partition tolerance \cite{Brewer2000}.

Consistency is a property of the technology platform and {\bf does not apply where the implementation does not use a distributed system}.  It is therefore preferable to define consistency in an implementation-independent manner, and to define it as an implementation constraint (which could be trivially observed by a non-distributed platform).

It might be important to define which objects are subject to a consistency constraint and which are not.  For example, perhaps objects that are not shared between participants need not be subject to the consistency constraint?  This would require the ISDA CDM to distinguish between ``private'' objects and ``shared'' objects, and opens up the broader discussion of ``shared with whom?'' --- for example, there might be three levels of sharing such as ``private'', ``counterparties for this trade'', and ``everyone''.  Or there might be more such levels.

It is likely that the ISDA CDM definition of consistency would {\em not} be predicated on the type of object being shared --- i.e. it is likely that the ISDA CDM would not wish to constrain the sharing of any particular type of object, but would specify that where any objects are shared the multiple copies must be consistent.
For example it may be useful to be able to share not only data but also code (which might conceivably be source code, byte code, object code or binary code).

\section{States, state transitions and events}
This section discusses the ISDA CDM definition as a state transition system, including the meaning of states, state transitions, events and operations.  The terminology of \cite{ISDA-CDM} is not followed exactly, since the aim is to provide a more detailed discussion.

Since the model will contain a variety of names, both names of types and names of functions, a naming convention is used from now on: names of user-defined types will end in $\_t$ whereas names of functions and other identifiers such as function parameter names will not.  
Tag names will always start with an uppercase letter.

\subsection{Operations, events and state}
This section starts with the concept of an $operation$, which is defined herein to be two ordered sequences of events, known as ``Before'' events and ``After'' events.  Thus, the type of all operations is given by $operation\_t$ defined below (where the first list $[event\_t]$ holds the ``Before'' events and the second list $[event\_t]$ holds the ``After'' events):
 \vspace{11pt}

\begin{tabular}{ll}
$operation\_t $&$::= \;Operation\;[event\_t]\;[event\_t]$\\
\end{tabular}
\vspace{11pt}

\noindent
This requires a definition of the type of events, called $event\_t$, which follows the format used in the examples in \cite{ISDA-CDM}:
\vspace{11pt}

\begin{tabular}{lll}
$event\_t $&$::= \;NoEvent $&$| \;Event \;eventid\_t \;party\_t \;party\_t \;amount\_t \;econid\_t$\\
$eventid\_t $&$::= \;NoEventID $&$| \;EventID \;num$\\
$party\_t $&$::= \;NoParty $&$| \;PartyID \;num$\\
$amount\_t $&$::= \;NoAmount $&$|\;Amount\;num$\\
$econid\_t $&$::= \;NoEconomics $&$|\; EconomicsReference \;num$
\end{tabular}
\vspace{11pt}

\noindent
Thus, in the above specification, an event is either the tag $NoEvent$ or the tag $Event$ followed by:
\begin{itemize}
\item
an event identifier (that is either the tag $NoEventID$ or the tag $EventID$ followed by a number), 
\item
two party identifiers (each is either the tag $NoParty$ or the tag $PartyID$ followed by a number), 
\item
a quantity (that is either the tag $NoAmount$ or the tag $Amount$ followed by a number), and 
\item
an economic description (which is either the tag $NoEconomics$ or the tag $EconomicsReference$ followed by a numeric reference). 
\end{itemize} 

\noindent
In \cite{ISDA-CDM} there is also mention of a smaller aggregation of data that is the same as the $Event$ tag of $event\_t$ but without the event identifier; this smaller aggregation is called a ``state'' but this term is misleading since ``state'' is normally a static property (e.g. of the contract as a whole) whereas this aggregation of data describes a dynamic action to change state.\footnote{It is assumed that a function $statetransition$ exists that takes an operation and the current state of the contract (let it be available inside the economic description of type $econid\_t$) and produces a new state of the contract.}

An ``event'' is an object of type $event\_t$ and it specifies an action of some kind that materially alters the state of the contract, where the term ``state of the contract'' is used in this paper to mean a static description of all of the information relating to the contract at a given time.

Thus an ``event'' is a specification (using a standard aggregation of information items) of a state transition.  However, the specification does not give precise before and after definitions of the state of the contract.  Rather, it gives an abstract specification of an action where a thing of value is passed from one party to another (which could presumably be payment or delivery) or where there is a gross change to the economic description.  This is somewhat unusual, but is motivated by the desire to define a common design pattern that can be reused at different levels.\footnote{It is not yet known whether this is the ``ideal'' design pattern for this purpose, nor yet is it known how to distinguish a good design pattern from a bad design pattern for this purpose.  In \cite{ISDA-CDM} the concept of a design pattern repeated at different levels is given the name ``fractal symmetry'', but that term is avoided in this paper.}

What do the items of an ``event'' mean?  We provide suggestions below:
\begin{description}
\item[$eventid\_t$]
This is the type of event identifiers (assumed in this model to be numeric, but could for example be alphanumeric); the assumption here is that every event has a separate identity and that, for example, event identifiers could be captured in a detailed provenance history.
\item[$party\_t$]
Two parties are specified.  Despite a somewhat confusing discussion in the definition, it is assumed that the ordering of these two parties in the event data structure is important: during a transfer of value, one will be the sending party and the other will be the receiving party.  For the purposes of this model, it is assumed that all parties to the contract (or no party) can be identified either using the tag $NoParty$ or a tag $PartyID$ followed by a numeric identifier --- for example, the economic description might hold a table of parties, and the numeric identifier would be an index into this list.  It is also assumed that the first party is the sending party and the second is the receiving party.
\item[$amount\_t$]
A single quantity is provided, and this is a numeric value to specify how many units of value are transferred.  
\item[$econid\_t$]
A single reference is provided to the ``economic description'' of the contract; in this simple model it is either the tag $NoEconomics$ or the tag $EconomicsReference$ followed by a numeric reference, but in general terms the reference need not be numeric and could for example be alphanumeric so that an alphanumeric cryptographic hash could be used.
\end{description}

\noindent
Note that an event only specifies a bilateral transfer; thus multilateral transfers must be split into a sequence of bilateral transfers.

\subsection*{Issues}
Several issues arise from the specifications of $operation\_t$ and $event\_t$ given above,  deriving directly from the draft ISDA CDM definitions:
\begin{itemize}
\item
The semantics of ``Before Events'' and ``After Events'' are not clearly defined.  Are ``Before Events'' intended to be actioned as part of an operation?  or must they ``have been actioned'' in the past, and if the latter how long ago must these events have been actioned?  Or are they not meant to be actioned but instead act merely as placeholders for useful contextual information?
\item
The units for $amount\_t$ are context-dependent and must be defined elsewhere --- with the current model, the only place where such details can be provided is in the economic description, but within one contract there may be transfers of value with varying units.  
\item
It appears not to be possible to determine from the event data whether an action is a delivery or a payment.  For simple contracts it might be possible to deduce direction rules from the economic description (e.g. transfers from party 1 are always deliveries and transfers from party 2 are always payments) but in general it will not be that simple.
\end{itemize}

\subsection{Example operations}
The draft ISDA CDM definition \cite{ISDA-CDM} gives numerous examples of operations, a few of which are repeated below as specifications.  In each case a function is specified that, when applied to appropriate arguments, will generate the data of type $operation\_t$ that gives exactly the sequence of ``Before'' events and ``After'' events defined for the relevant operation in \cite{ISDA-CDM}:\footnote{This approach is taken to clarify precisely what information is required for each operation.}

\begin{description}
\item[New]
In \cite{ISDA-CDM} this is illustrated using an empty ``Before'' event and a single ``After'' event.  
This can be modelled as a function that is parameterised on the event identifier and the reference to the new economic description.  

\begin{tabular}{lll}
$new \;(id, \;e, \;party1, \;party2, \;quantity) $\\
$\;\;\;\;\;\;\;\;\;\;\;\;\;\;\;\;\;\;\;\;\;\;\;\;= \;Operation\; [NoEvent]\; [(id, \;party1, \;party2, \;quantity, \;e)]$
\end{tabular}
\vspace{11pt}

\noindent
It would appear to be equally valid to use the following alternative definition (since there is no information in the ``Before'' event to be linked to the ``After'' event):

\begin{tabular}{lll}
$new \;(id, \;e, \;party1, \;party2, \;quantity) $\\
$\;\;\;\;\;\;\;\;\;\;\;\;\;\;\;\;\;\;\;\;\;\;\;\;= \;Operation\; []\; [(id, \;party1, \;party2, \;quantity, \;e)]$
\end{tabular}
\vspace{11pt}

\item[Terminate for cash]
In this example the ``Before'' event is absolutely necessary because it appears to be the only place where the economic descriptor is available for the contract that will be terminated. However the previously noted issue remains: what exactly is the meaning of a ``Before'' event?
This can be modelled as a function that is parameterised on all the required data for the event description.
\vspace{11pt}

\begin{tabular}{lll}
$terminate\_for\_cash $&$(id1,\;id2, \;from, \;to, \;quantity, \;cashamount, \;e) $\\
                      $\;\;\;\;\;\;\;\;\;\;\;\;\;\;\;\;= \;Operation $&$[(id1, \;from, \;to, \;quantity, \;e)]$\\ 
                         &$[(id2, \;from, \;to, \;cashamount, \;NoEconomics)]$\\
\end{tabular}
\vspace{11pt}

\noindent
Here the tag $NoEconomics$ is preferred to the description ``Cash'' used in \cite{ISDA-CDM}.  Furthermore, although it may not be clear from the syntax, there is a semantic dependency such that it is the contract $e$ in the ``Before'' event that is replaced by $NoEconomics$ in the ``After'' event.
\item[Amend]
This is intended to be an update to the economic description (\cite{ISDA-CDM} mentions that there should perhaps be constraints on which updates are permissible).  Again, there is a need to specify {\em which} contract is being amended, and the ISDA CDM draft definition uses a ``Before'' event to provide this information:
\vspace{11pt}

\begin{tabular}{ll}
$amend \;(id1,\;id2,\; from, \;to, \;quantity1, \;quantity2, \; e1,\; e2)$\\
$\;\;\;\;\;\;\;\;\;\;\;\;\;\;\;= \;Operation\;\;\;\;[(id1, \;from, \;to, \;quantity1, \;e1)]$ \\
$\;\;\;\;\;\;\;\;\;\;\;\;\;\;\;\;\;\;\;\;\;\;\;\;\;\;\;\;\;\;\;\;\;\;\;\;\;\;\;\;[(id2, \;from, \;to, \;quantity2, \;e2)]\;\;\;\;$
\end{tabular}
\vspace{11pt}

\noindent
In the above specification, economic description $e1$ is being replaced by economic description $e2$.  The definition in \cite{ISDA-CDM} also shows the quantity being modified.

\item[Split]
This operation has a complex dependency on previously existing information.  A direct specification taken from the example in \cite{ISDA-CDM} gives (where $q1+q2=Q$):
\vspace{11pt}

\begin{tabular}{lll}
$split \;(id1,\;id2,\; from,\;to,\;t1,\;t2,\;q1,\;q2,\;e)$\\
$\;\;\;\;\;\;\;\;\;\;\;\;\;\;\;\;\;\;\;\;\;\;\;\;\;\;\;\;\;\;\;\;\;\;\;\;\;\;= \;Operation$     &$[(id1, \;from, \;to, \;q1+q2,$  &$e)]$ \\
                                                                                             &$ [(id2, \;from, \;t1, \;q1,$     &$e),$\\
                                                                                             &$\;(id2, \;from, \;t2, \;q2,$     &$e)\;\;]\;\;$
\end{tabular}
\vspace{11pt}

\noindent
However, several issues arise from this specification:
\begin{itemize}
\item
The amount $q1+q2$ is said to be ``the original trade quantity''.  This places a further constraint on the ``Before'' event: not only must it be an event relating to the contract that must be split (so that the economics descriptor $e$ is available) but also it must be {\em the event} that contained the original trade quantity.  This implicitly constrains the ``Before'' event to be the immediately preceding event, but without a rigorous semantic definition of these operations, several different implementations could be imagined.
\item
The multiple ``After'' events all have the same Event Identifier ``id2''.  Should this therefore be renamed an ``Operation Identifier''?
\item
Is the ``Before'' event a delivery or a payment?  If $q1$ and $q2$ are splits across funds then they are presumably amounts of money, and if $q1+q2=Q$ then presumably the ``Before'' event must have been a transfer of money, but the semantics are not at all clear.
\item
If the economic description contains the state of the contract, then the economic description ID should change after each event (since the state will have changed) yet in the example for ``split'' in \cite{ISDA-CDM} the economic description ID remains the same; is this simply an error in the draft definition, or is there a reason why it must remain the same?
\end{itemize}

\noindent
The issues of (i) the semantics of ``Before'' events, (ii) distinguishing deliveries from payments, and (iii) clearly expressing how contract state changes, appear to be fundamental and require further analysis.

\item[Partial Assignment]
Partial Assignment exposes similar issues to Split.  It can be modelled in a very similar way. A direct specification taken from the example in \cite{ISDA-CDM} gives:
\vspace{11pt}

\begin{tabular}{lll}
$partial\_assign \;(id1,\;id2,\; from,\;to,\;t1,\;q1,\;q2,\;cash,\;e1,\;e2)$
\end{tabular}
\begin{tabular}{lll}
$\;\;\;\;\;\;\;\;\;\;\;\;\;\;\;\;= Operation\;$&$[(id1, \;from, \;to, \;q1+q2,$&$e1)]$ \\
                 &$ [(id2, \;from, \;to, \;q1,$&$e2),\;\;\;\;$\\
                 &$ \;(id2, \;from, \;t1, \;q2,$&$e2),\;\;\;\;$\\
                 &$ \;(id2, \;to, \;t1, \;cash,$&$NoEconomics)\;\;]\;\;$
\end{tabular}
\vspace{11pt}

\noindent
Reading the above specification it is not clear whether $NoEconomics$ replaces the original contract $e1$ or the contract $e2$, nor whether $e2$ replaces $e1$ before then being replaced by $NoEconomics$.  Clearly, the intention is that an operation is able to manipulate more than one contract at a time, but the draft definition lacks a rigorous semantics for what a sequence of events means where there are multiple contracts.  This appears to be a substantial problem with the current definition of events and operations and requires further analysis.

\item[Tear-up]
The ``Tear-up'' example of \cite{ISDA-CDM} is unusual in that it has four ``Before'' events and five ``After'' events.
A direct specification taken from the example in \cite{ISDA-CDM} gives (where {\em qx} and {\em ex} represent a potentially large number of contracts that are cancelled)
\vspace{11pt}

\begin{tabular}{lll}
$tear\_up\;(id,\; from,\;to,\;qx,\;q2,\;q3,\;q4,\;q5,\;x,\;y,\;cash,\;ex,\;e2,\;e3,\;e4,\;e5,\;e6)$
\end{tabular}
\begin{tabular}{lll}
$\;\;\;\;\;\;\;\;\;\;\;\;\;\;\;\;= Operation\;$&$[(id, \;from, \;to, \;qx,$&$ex),$ \\
                 &$ \;(id, \;from, \;to, \;q2,$&$e2),\;\;\;\;$\\
                 &$ \;(id, \;from, \;to, \;q3,$&$e3),\;\;\;\;$\\
                 &$ \;(id, \;from, \;to, \;q4,$&$e4)]\;\;\;\;$\\                
                 &$ [(id, \;from, \;to, \;q2,$&$e2),\;\;\;\;$\\             
                 &$ \;(id, \;from, \;to, \;q3-x,$&$e3),\;\;\;\;$\\                 
                 &$ \;(id, \;from, \;to, \;q4+y,$&$e4),\;\;\;\;$\\                 
                 &$ \;(id, \;from, \;to, \;q5,$&$e5),\;\;\;\;$\\                 
                 &$ \;(id, \;from, \;to, \;cash,$&$e6)\;\;]\;\;$
\end{tabular}
\vspace{11pt}

\noindent
Issues that arise from the example in \cite{ISDA-CDM} include:
\begin{itemize}
\item
Should all the ``Before'' events have the same event ID as all the ``After'' events? (and, as before, should this be renamed an ``operation ID''?)
\item
The first ``After'' event appears to be a transfer of value from one party to another, yet it is commented in \cite{ISDA-CDM} as being ``no change''.  This is confusing; preferably, if no change is required then there would be no event.
\item
The fourth ``After'' event is commented in \cite{ISDA-CDM} as being ``new'', 
presumably meaning a new contract, yet without that comment there is nothing to indicate that a new contract is being created (rather than one of the existing contracts being updated with new economics).  
The semantics relating to contract creation and contract replacement (or state transition) are opaque; they must be clarified and made rigorous.
\item
The fifth ``After'' event in the above specification indicates a new economic description $e6$, whereas in the example in \cite{ISDA-CDM} it is given the name ``CCY'' (presumably meaning ``Currency'').  In other examples in \cite{ISDA-CDM} the name ``Cash'' is used for the economic description where a fee is paid, which in the other specifications given above has been renamed as $NoEconomics$, yet the use of the name ``CCY'' implies that this case is different (which is why $NoEconomics$ has not been used here).
\end{itemize}

\end{description}

\noindent
The attempt to model just a few of the examples from \cite{ISDA-CDM} has exposed several key issues that require further discussion.  As other examples of events (such as output events where the smart contract code needs to attract attention from a human operator) are explored in detail, different issues may also arise.  It appears likely that each new example operation might expose a different problem, and without a rigorous definition it is not possible to make a rigorous analysis.  As a result it will be difficult to achieve confidence that all possibilities have been considered.  This requires substantial further analysis and may require modification to the way in which operations, events and states are defined and processed.

\section{Summary and Conclusions}
The ISDA CDM draft definition is a complex document, with definitions in a mathematical style appearing almost from the start.  This paper has therefore sought to provide (i) 
commentary, discussion and feedback, (ii)
rigour and additional thinking to contribute to improvements in the next iteration of the ISDA CDM definition,
(iii)
support for people who are looking to understand the draft definition,  and (iv)
support for people aiming to undertake design and prototyping of the ISDA-CDM in software.  

To provide the most value initial focus has been given to the foundational topics of data lineage,
timestamps, consistency of replicated data, operations, events, state and state transition.  During analysis and discussion of those topics, several key issues have been exposed and these are summarised below:

\begin{itemize}
\item
It is not yet clear whether the draft definition is too generic or insufficiently generic, nor any agreed way to decide what level of genericity would be ``best''.
\item
Timestamps, ordering of events, and consistency are normally straightforward for non-distributed implementations, but for distributed implementations (including DLT) they can be problematic. Since the ISDA CDM aims to be implementable on distributed as well as centralised platforms further thought must be given to whether constraints or guidelines should be incorporated into the ISDA CDM definition.  
\item
Operations are defined as a sequence of ``Before'' events followed by a sequence of `After'' events, yet the semantics of this definition are opaque.  It is not stated whether ``Before'' events have or have not yet been actioned, and if the former whether they must be immediately preceding actions or have been actioned at any time in the past.  Furthermore, there is complex unstated semantic linkage between names used in ``Before'' events and the semantics of the ``After'' events (such as where a contract is partially assigned, and in general where a contract is created or replaced). This requires considerable further analysis.
\item 
Events are state transitions (they change the state of the contract) yet their definitions do not include components of the contract state.  Instead, they are given a more abstract definition.  This provides a design pattern that can be used at different levels, yet it obscures the semantics of state transition.  Such a design pattern might be advantageous if a rigorous specification could be provided; however, the lack of rigour is currently problematic in several ways and therefore perhaps the definitions of events and operations need to be reconsidered.
\item
Where an operation must manipulate more than one contract, a rigorous syntax and semantics to describe these manipulations, including sequences of events each of which might correspond to a different contract, is currently lacking.  This requires substantial further analysis.
\item
Each new example operation has exposed slightly different problems, and the absence of rigorous definitions makes it impossible to provide analysis to ensure properties such as completeness, correctness, and consistency.
This requires substantial further analysis and may require modification to the way in which operations, events and states are defined and processed.
\item
At a slightly lower level of detail:
\begin{itemize}
\item
The direction of a transfer of value within an event is not clearly specified, nor whether it is a delivery or payment, nor what are the units for the stated quantity.
\item
If an event changes the state of a contract, and if the state of a contract is contained within the economic description, then should the economic description not change after every event?
\item
Events are dynamic objects (they effect a change) and are specified as a collection of information called a ``state''.  This terminology is unfortunate, since the term ``state'' is normally used to refer to a static property of a system: e.g. the current state of a contract.
\end{itemize}
\end{itemize}

\noindent
The ISDA CDM draft definition is at an early ``discussion'' stage and this paper provides only an initial analysis and discussion.  There is substantial further work to be done: as stated in the introduction, it is hoped that the points raised in this paper will stimulate debate and encourage feedback so that this working paper can be improved and extended.

\section*{Acknowledgements} 
The author would like to thank Lee Braine (Barclays) and Rajagopalan Siddharthan (Barclays) for their helpful feedback on drafts of this paper. 
Thanks are also due to Clive Ansell (ISDA), Ian Sloyan (ISDA) and Tim Smith (Credit Suisse) for many detailed discussions relating to the ISDA CDM.

\pagebreak
\section*{Appendix: features and benefits of the ISDA CDM}

As the ISDA CDM definition moves towards its next iteration, there will be many design decisions to be made and at each decision point it will be helpful to have a set of agreed principles by which to determine which is the better of two choices.  One approach would be to repurpose a list of benefits as a list of objectives.  These might further be ordered in terms of importance, and might be linked to specific features in the current design.

From \cite{ISDA-CDM-2} we summarise the following motivating benefits, with links to specific features that are given in a following list.  These benefits are not ordered: ranking in terms of importance is left for future work.

\subsection*{ISDA CDM benefits}
\begin{itemize}
\item
Facilitate uptake of opportunities for improved process transformation (from features F2 and F3).
\item
Remove substantial
human involvement in standard processes (from F1, F2 and F3).
\item
Facilitate improved scalability of processing infrastructure (from F4).
\item
Reduce the current duplication and fragmentation of effort and information (from F5).
\item
Facilitate standardised and enhanced analysis and risk management  processes (from F6).
\item
Provide enhanced trade processing capabilities (from F4 and B4).
\item
Provide greater transparency between market participators and regulators (from F7).
\item
Reduce operating expense (from F4).
\item
Reduce the cost of compliance with regulatory requirements (from F7).
\item
Provide regulatory reports of greater quality, accuracy and timeliness (from F7).
\item
Lower the barrier to entry for new financial technology service providers (from F4).
\item
Promote innovation via facilitation of uptake of new technologies (from F1, F2 and F3).
\end{itemize}

\subsection*{ISDA CDM features}
\begin{itemize}
\item
F1: Reduced reliance on legacy technology platforms.
\item
F2: Support for the running of autonomous code such as smart contracts \cite{szabo1997,stark2016,SCT2016}.
\item
F3: Support for 
new technology platforms such as distributed ledger technology (DLT).
\item
F4: Simplification and non-differentiated standardisation of technology infrastructure.
\item
F5: Provision of an authoritative central record --- a single agreed and trusted source of information in a standardised form.
\item
F6: Improved data capture, e.g. automatic capture of data lineage  in a standard format.
\item
F7: Provision of detailed, standardised and potentially automated reporting (including compliance reporting to regulatory agencies).
\item
F8: A consistent hierarchical representation across trades, portfolios and events.
\end{itemize}

\pretolerance=-1
\tolerance=-1
\emergencystretch=0pt

\pagebreak
\bibliography{CDM}

\begin{thebibliography}{10}

\bibitem{Brewer2000}
E.~Brewer.
\newblock {T}owards {R}obust {D}istributed {S}ystems.
\newblock In {\em Proc. 19th Ann. ACM Symp. Principles of Distributed Computing
  (PODC 00)}, pages 7--10. ACM, 2000.

\bibitem{SCT2016}
{C. D. Clack}, {V. A. Bakshi}, and {L. Braine}.
\newblock {{Smart Contract Templates: foundations, design landscape and
  research directions}}.
\newblock {\em The Computing Research Repository (CoRR)}, abs/1608.00771, 2016.
\newblock Also at arXiv.org: \url{http://arxiv.org/abs/1608.00771/}.

\bibitem{Clack1995}
C.~D. Clack, C.~Myers, and E.~Poon.
\newblock {\em {P}rogramming with {M}iranda}.
\newblock Prentice Hall, 1995.

\bibitem{Fidge1988}
C.~J. Fidge.
\newblock {T}imestamps in {M}essage-{P}assing {S}ystems {T}hat {P}reserve the
  {P}artial {O}rdering.
\newblock In {\em Proc. of the 11th Australian Computer Science Conference
  (ACSC'88)}, pages 56--66, 1988.

\bibitem{ISDA-CDM-2}
{I}nternational {S}waps and {D}erivatives {A}ssociation.
\newblock {ISDA} {CDM} {D}ata and {P}rocess {S}tandards, September 2017.
\newblock
  \url{https://www.isda.org/a/0EKDE/isda-mitoc-data-and-process-standards-initiative-standard-outreach-dec.pdf}.

\bibitem{ISDA-CDM}
{I}nternational {S}waps and {D}erivatives {A}ssociation.
\newblock {ISDA} {C}ommon {D}omain {M}odel {V}ersion 1.0 {D}esign {D}efinition
  {D}ocument, 2017.
\newblock \url{http://assets.isda.org/media/edbc86cb/abd4def0-pdf/}.

\bibitem{Lamport1978}
L.~Lamport.
\newblock {T}ime, clocks, and the ordering of events in a distributed system.
\newblock {\em Communications of the ACM}, 21(7):558--565, 1978.

\bibitem{Mattern1988}
F.~Mattern.
\newblock {V}irtual {T}ime and {G}lobal {S}tates of {D}istributed {S}ystems.
\newblock In {\em Proc. Workshop on Parallel and Distributed Algorithms}, pages
  215--226. Elsevier, 1988.

\bibitem{stark2016}
J.~Stark.
\newblock Making sense of blockchain smart contracts, 2016.
\newblock \url{http://www.coindesk.com/making-sense-smart-contracts/}.

\bibitem{szabo1997}
N.~Szabo.
\newblock {F}ormalizing and {S}ecuring {R}elationships on {P}ublic {N}etworks.
\newblock {\em First mind}, 2(9), 1997.

\bibitem{Thompson1995}
S.~Thompson.
\newblock {\em {M}iranda: the {C}raft of {F}unctional {P}rogramming}.
\newblock Addison-Wesley, 1995.

\bibitem{Turner-executablespec}
D.~A. Turner.
\newblock {F}unctional {P}rograms as {E}xecutable {S}pecifications.
\newblock In {\em Mathematical Logic and Programming Languages}, pages 29--54.
  Prentice Hall, 1985.

\bibitem{Turner1986}
D.~A. Turner.
\newblock {A}n {O}verview of {M}iranda.
\newblock {\em SIGPLAN Notices}, 21(12):158--166, 1986.

\end{thebibliography}
\bibliographystyle{plain}

\end{document}